\documentstyle[preprint,aps,floats]{revtex}
\input epsf %
\begin{document}
\draft
\title{Experimental Status of Gravitational-Strength Forces in the Sub-Centimeter Regime}
\author{J. C. Long, H. W. Chan, and J. C. Price}
\address{Department of Physics - CML, University of Colorado, Boulder CO 80309 USA}
\date{\today}
\maketitle
\begin{abstract}
We review the experimental constraints on additional macroscopic
Yukawa forces for interaction ranges below 1 cm, and summarize several
theoretical predictions of new forces in this region.  An experiment
using 1 kHz mechanical oscillators as test masses should be sensitive
to much of the parameter space covered by the predictions.
\end{abstract}
% body of paper here

\narrowtext

\section{Introduction}

Experimental searches for a deviation from Newton's law of universal
gravitation have received a great deal of attention over the past
three decades.  Motivation has come from reports of experimental
anomalies, and from new theoretical predictions.  Searches have
involved a great variety of experiments on the planetary, lunar,
geological, and laboratory scales.  A recent review by Fischbach and
Talmadge \cite{Fischbach} shows that new forces with a strength weaker than or
comparable to gravity have been excluded over distances ranging
between 1 and $10^{17}$ cm.  Experiments have only marginally explored the
distance range under 1 cm, however, and there is little knowledge of
gravity itself in this range \cite{Price}.  Furthermore, the sub-centimeter
region is of increasing experimental interest, given a number of
recent predictions of new forces from modern theories which attempt to
unify gravity with the strong and electroweak interactions.

In this paper we present a review of the existing experimental limits
and some of the recent theoretical predictions.  We then describe an
experiment which we expect to be sensitive to gravity at distances as
short as 50 $\mu$m, and to much of the parameter space for the new predictions.
\section{Current Limits}

Experimental results from searches for new long-range forces of
gravitational strength have been parameterized in several ways \cite{Cook}.
The Yukawa interaction is most commonly used, and we adopt it here.
The potential due to gravity and an additional Yukawa interaction
between two macroscopic objects may be written
\begin{equation}
V(r) = -\int{dr_{1}} \int{dr_2} \frac{G\rho_{1}(r_{1})\rho_{2}(r_{2})}{r_{12}} [1 + \alpha \exp{(-r_{12}/\lambda)}],
\label{eq:FY}
\end{equation}
where $G$ is the Newtonian gravitational constant, $r_{12}$ is the distance
between $r_{1}$ and $r_{2}$, $\rho_{1}(r_{1})$ and $\rho_{2}(r_{2})$
are mass densities, $\alpha$ is the strength of the Yukawa
force relative to gravity, and $\lambda$ is the range.  If, as is likely, the
Yukawa forces couple to quantities other than mass, then $\alpha$ will have a
slight composition dependence \cite{Fujii}, which we do not consider.

The current experimental limits on $\alpha$, for Yukawa ranges between
1 $\mu$m and 1 cm, are illustrated in Fig.~\ref{fig:lim}, together 
with the limit anticipated from the experiment outlined in 
Sec.~\ref{sec:meas} below.  Also
indicated are specific predictions for new long-range forces,
described in Sec.~\ref{sec:theory}.  The region above and to the
right of the  bold curves is excluded by experiments falling into two
general categories: classical gravity measurements and Casimir force
measurements.

\subsection{Classical Gravity Experiments}

For values of $\lambda$ greater than 100$\mu$m, the most stringent
limits on $\alpha$ can be obtained from torsion balance experiments
designed to test the gravitational inverse-square law.  Well before
the interest in gravitational measurements brought on by suggestions
of a ``fifth force,'' the report of a possible deviation in a
measurement by Long \cite{Long} motivated several similar experimental
programs (see, for example, Refs. \cite{Chan,Ogawa,Chen}) which 
have apparently ruled out any evidence for the anomaly \cite{Michaelis}.

One such experiment is the Irvine 2-5 cm null measurement
\cite{Hoskins}, which yields the best limits in the range near 1 cm.
This experiment used a torsion balance to measure the force on a test
mass suspended inside a hollow cylinder, a non-zero result for which
would have been evidence for an anomalous force.  After an exhaustive
consideration of possible systematic effects, and careful analysis of
inhomogeneity in the cylinder, the authors derived the limit curve
reproduced in Fig.~\ref{fig:lim}, corresponding to a force
sensitivity of about $4 \times 10^{-10}$ dyne.  Only the curve
corresponding to the (more sensitive) limiting case of a hypothesized
attractive Yukawa force ($\alpha > 0$) is shown. 

Below the range of a few millimeters, the most stringent limits
may be obtained from the torsion balance experiment of Mitrofanov and
Ponomareva \cite{Mitrofanov}.  This experiment used a dynamic-resonant
technique to distinguish the gravitational interaction of the test
masses from backgrounds.  A test mass, initially spaced 3.8 mm
perpendicular to one end of the balance, was moved synchronously with
the oscillations of the balance.  The authors were able to achieve a
sensitivity of about $3 \times 10^{-10}$ dyne, for integration times
of about $10^{3}$ s.  The ratio of the forces measured at 3.8 mm and
6.5 mm was used to obtain a curve similar to that shown in
Fig.~\ref{fig:lim}.

We have re-analyzed the data of this experiment, using Eq. 7 of 
Ref.~\ref{ref:Mit}, to extend the limit to shorter distances than
those considered by the authors.  The resulting curve is shown in 
Fig.~\ref{fig:lim}.  The 1$\sigma$ errors on the test mass separations, 
apparently not taken into account in the original report, are added in
quadrature to the 1$\sigma$ errors on the measured force ratio, so
that the limit we derive is slightly less sensitive than that reported
in Ref.~\ref{ref:Mit}.  We have not illustrated the cases for the 
separate hypotheses of attractive and repulsive forces, which have
little effect on the results.

Other inverse-square law and Cavendish experiments have been performed
with inter-mass separations on the order of 1 cm, including that of
the Tokyo group \cite{Mio}.  None, however, reach stronger limits in
the range below 1 cm.  The Rot-Wash experiment, which uses a 2620 kg 
depleted uranium attractor rotated around a torsion balance with lead
and copper test masses, yields a limit nearly competitive with the
Irvine result at 1 cm \cite{Gundlach}.  This assumes a Yukawa force
which couples to mass; the use of uranium makes this experiment
especially sensitive to composition-dependent forces.  Assuming a
coupling to $N - Z$, the Rot-Wash experiment has set a limit 10 times
more stringent than the Irvine experiment at 1 cm.

\subsection{Casimir Force Experiments}

In 1948, H. G. B. Casimir showed that a weak attractive force should
be present between a set of flat, parallel conducting plates, due to
the zero-point fluctuations of the electromagnetic field
\cite{Casimir}.  This force is given by
\begin{equation}
F_{c} = \frac{\pi^{2}}{240} \frac{\hbar c}{r^{4}} A,
\label{eq:cas}
\end{equation}
where $r$ is the separation and $A$ is the area.  A similar effect,
called the retarded Van der Waals force, occurs with dielectric plates
and is given by an expression which differs from Eq.~\ref{eq:cas} only
by a numerical factor of order unity \cite{Chu}.

For a given value of $\lambda $, the most stringent limits 
on $\alpha$ can be obtained from Casimir and Van der Waals force 
data taken with mass 
separations of order $\lambda $, given the strong dependence of 
Eq.~\ref{eq:cas} on $r$.  These experiments usually investigate
separation distances well below 1 mm.  In practice, the best limits to
date are derived from measurements of the force between a flat plate
and a spherical shell, which varies as $r^{-3}$.  The experiment by 
Derjaguin, Abrikosova, and Lifshitz \cite{Derjaguin} done in 1956
provided the most sensitive limits in the range less than 100 $\mu$m 
for 40 years, until the recent measurement by Lamoreaux
\cite{Lamoreaux}.

This experiment measured the attraction between a flat, metal-plated 
quartz disk and a metal-plated spherical lens separated by distances 
ranging from about 0.5 to 10 $\mu$m.  Using data from this experiment,
we have derived the limits presented in Fig.~\ref{fig:lim}.  For
several values of $\lambda $ in the range 1-100 $\mu$m, a
least-squares fit is made with the data in the measured force plot 
(Fig. 3, top, of Ref.~\ref{ref:Lam}), to the expression
\begin{equation}
F^{m}_{i} = F^{T}_{c} (a_{i} + a_{0}) + \frac{\beta}{a_{i} + a_{0}} 
+ b + F_{Y}[\alpha, \lambda, \rho_{j}, d_{j}, (a_{i} + a_{0})].
\label{eq:lam}
\end{equation}
This is based on Eq. 7 of Ref.~\ref{ref:Lam}, where the first three 
terms are the Casimir, electromagnetic, and constant terms used by 
Lamoreaux to fit the data (with the same fit parameters $a_{0}$, 
$\beta$, and $b$).  We have added a Yukawa term $F_{Y}[\alpha,
\lambda, \rho_{j}, d_{j}, (a_{i} + a_{0})]$, in which $\alpha$ is 
taken as an additional fit parameter.  The Yukawa term is a function 
of the densities $\rho_{j}$ and thicknesses $d_{j}$ of the gold,
copper, and quartz comprising the Casimir test plates
\cite{Lamoreaux2}.

For each value of $\lambda$ investigated, $\alpha$ is then fixed at 
various values within a range of the fit value, and the fit is 
performed again.  The 1$\sigma$ acceptance region in $\alpha$ is then 
taken to be all the points in the ($\alpha, \lambda$) parameter space 
which have a $\chi^{2}$/d.o.f. within 1.0 of the minimum value.  
This region, bounded above by the curve in Fig.~\ref{fig:lim}, is 
roughly one order of magnitude more stringent than the Derjaguin 
limit over the entire range shown, for positive values of $\alpha$ 
\cite{Lamoreaux3}.  It is very important to note that the measured 
force plot presumably represents less than one percent of the total 
data set \cite{Lamoreaux4}.  It is therefore likely that the limits 
attainable from a complete re-analysis of this experiment are at 
least another order of magnitude more sensitive than those presented 
in Fig.~\ref{fig:lim}.

\subsection{Other Limits}

Limits on new forces in the range considered here have been obtained 
in experiments on other systems, but they are not sensitive enough 
to compete with the limits shown in Fig.~\ref{fig:lim}.  Bracci, 
Fiorentini, and Tripiccione \cite{Bracci} set limits on long-range 
nucleon-nucleon interactions from a re-analysis of a variety of 
experiments.  In addition to macroscopic systems, they consider 
the spectroscopy of hydrogenic and mesic molecules, antiprotonic 
atoms, and data from low energy nucleon-lead and nucleon-nucleon 
scattering experiments.  The limits on the coupling strength derived 
from these elementary systems would have to improve by many orders 
of magnitude in order to appear in Fig.~\ref{fig:lim}.  Leeb and 
Schmiedmayer have derived similar constraints from neutron scattering 
experiments below 10 keV, and from neutron optics experiments
\cite{Leeb}.  Studying the difference between scattering lengths 
measured in these two classes of experiments, they extract limits on 
the size and range of an additional Yukawa term in the neutron-nucleus
interaction.  While more sensitive than those obtained in
Ref.~\ref{ref:Bra}, these limits are again many orders of magnitude 
too weak to appear in Fig.~\ref{fig:lim}.

In general, since the electrostatic force between protons is $10^{36}$
times stronger than the gravitational force, no experiment on an 
elementary system which is dominated by electromagnetic effects can be
expected to provide limits strong enough in this range to compete 
with macroscopic measurements \cite{Barshay}.

\section{Theoretical Predictions}
\label{sec:theory}

As indicated in Fig.~\ref{fig:lim}, current experimental limits allow 
for the existence of forces many orders of magnitude stronger than 
gravity in the range below one millimeter.  Predictions of new
forces---usually involving very light bosons---have arisen 
in many contexts, from the early notion of Lee and Yang of a field 
coupled to baryon number \cite{Lee}, to modern ideas connected with 
string theory.  For experiments below one millimeter, several well-motivated
predictions can be addressed.

Many modern predictions arise in theories which attempt to describe 
gravitation and the other fundamental interactions in a single 
framework.  The mass of the light boson is typically given by \cite{Fujii}:
\begin{equation}
m \approx M_{P} \left( \frac{M}{M_{P}} \right)^{n}
\label{eq:boson}
\end{equation}
where $M_{P} \approx 10^{19}$ GeV is the Planck mass (or string scale),
$M$ is a mass scale connected with the known elementary particles, and
$n$ is a small integer.  For $n = 2$, the choice of $M = 10$ TeV
yields $m = 10^{-2}$ eV, corresponding to a Yukawa force with a range 
of $\lambda = 20 \mu$m.

An interesting class of predictions arises in superstring theories 
in which supersymmetry breaking occurs at low energies (less than 
about 100 TeV).  These theories are of interest as a way to address 
the supersymmetry flavor problem.  Superstring models generally 
contain gravitationally-coupled scalars called moduli which are 
massless at the string scale.  Dimopoulos and Giudice consider a 
``gauge-mediated'' model where the moduli acquire mass only from 
supersymmetry breaking, which they suppose to occur at 10-100 TeV 
\cite{Dimopoulos}.  This corresponds to Eq.~\ref{eq:boson} with 
$M$ = 10-100 TeV and $n = 2$.  The predicted strength and range of 
the modulus forces associated with the down quark, strange quark, 
and the gluon are shown in Fig.~\ref{fig:lim}.  Each prediction 
covers a region in the ($\alpha,\lambda$) plane because of 
uncertainties in the model parameters.  The predicted strengths are 
larger than gravity by several orders of magnitude, and the ranges 
are several orders larger than would be expected from
Eq.~\ref{eq:boson}.  Similar scenarios have been discussed 
previously by Banks, Berkooz, and Steinhardt \cite{Banks}.

Antoniadis, Dimopoulos, and Dvali have presented a related theory in 
which supersymmetry breaking occurs via Scherk-Schwarz
compactification at the weak scale ($M \approx$ 1 TeV)
\cite{Antoniadis}.  The authors 
show that theories involving this mechanism are likely to give rise 
to at least one scalar particle, the radius modulus, with a Compton 
wavelength corresponding to the size of the compact dimension.  In 
the model considered, the radius modulus has a Compton wavelength 
in the sub-millimeter range, and mediates a force with strength 1/3 
that of gravity, as shown in Fig~\ref{fig:lim}.

Other light scalars with stronger couplings are also possible.  The 
dilaton of string theory couples to matter with a strength roughly 
80 times that of gravity \cite{Taylor,Ellis}.  The mass of the dilaton
depends on unknown physics.  Dimopoulos and Giudice suggest a dilaton 
mass of about 10 eV for their scenario, but other authors take the 
dilaton mass to be an unknown parameter \cite{Gasperini}.  The
existing limits shown in Fig.~\ref{fig:lim} exclude a mass less than
$10^{-3}$ eV.
  
The axion is a light pseudoscalar boson motivated by the strong CP
problem of the standard model.  Laboratory experiments and
astrophysical bounds have left an allowed window for the axion mass 
of $10^{-6}$ to $10^{-3}$ eV, or $\lambda$ = 20 cm - 0.2 mm
\cite{Turner}.  However, the coupling $\alpha$ of the associated 
long-range force between unpolarized nucleons has been constrained, 
by recent measurements of the neutron electric dipole moment, to the 
region of the ($\alpha, \lambda$) plane shown in Fig.~\ref{fig:lim}.  
This is probably outside the reach of the experiment described in 
Sec.~\ref{sec:meas}, but could be a possible goal for the future 
\cite{Barbieri}.  Moody and Wilczek \cite{Moody} have argued that 
centimeter-scale experiments with polarized bodies would be more 
promising, and several attempts in this direction have been made 
\cite{Youdin,Ritter,Wineland,Venema}.

Still other models are motivated by cosmological observations.  
The oscillating-G cosmology of Steinhardt and Will involves a massive 
Brans-Dicke field which is left oscillating at high frequency after 
inflation \cite{Steinhardt}.  Yukawa corrections to the Newtonian 
potential are predicted, and useful bounds on a product of the
parameters of the model are found when these predictions are compared 
to experiments.  The experiment  described below would improve these 
bounds by a factor of 600 \cite{Will}.  A recent model due to Beane 
is based on more general assumptions.  The author argues that if all 
experimentally observable phenomena can be described in terms of 
effective field theories, then the observed smallness of the
cosmological constant implies the existence of new quanta
\cite{Beane}.  Furthermore, if a substantial component of the energy 
density of the universe in the present epoch is due to vacuum energy, 
then the range of the new quanta should be approximately 100 $\mu$m, 
as shown in Fig.~\ref{fig:lim}.  R. Sundrum develops these ideas 
further and proposes a specific mechanism that could provide a 
solution to the cosmological constant problem \cite{Sundrum}.  
In this theory, the new quanta are comprised of light gravitational 
strings.  The size of the cosmological constant is set by the scale 
of these strings (as opposed to the standard model scale), below which
the behavior of gravitational forces is expected to be quite
different.  Consistency with current cosmological and laboratory 
constraints suggests this scale to be between 10 $\mu$m and 1 cm.

Finally, we mention a theory which does not necessarily make recourse 
to new light particles.  In a recent discussion, Arkani-Hamed, 
Dimopoulos, and Dvali propose a theory in which gravitational and 
gauge interactions are united at the weak scale, which is taken to 
be the only fundamental short scale in nature \cite{Arkani}.  This 
model does not rely on supersymmetry and is of interest as a solution 
to the hierarchy problem.  The relative weakness of the 
gravitational interaction is a consequence of $n \geq 2$ new compact 
dimensions which are large compared to the weak scale, and in which 
gravitons propagate freely but the standard model fields do not.  
For the smallest number of new dimensions yet to be ruled out by 
experiment ($n = 2$), the size of the new dimensions is in the 
millimeter region.  Gravitational measurements in this regime would be
sensitive to potentials dictated by Gauss' law in ($4 + n$)
dimensions, or gravitational forces varying as $r^{-4}$.

From this discussion it is clear that forces of gravitational strength
are playing an increasingly important theoretical role, especially
since the advent of string theory.  There is ample motivation to push 
to the shortest ranges possible, where macroscopic experiments might 
provide a rare window on physics at the Planck scale.

\section{Short--Range Measurement with Mechanical Oscillators}
\label{sec:meas}

As the length scale $L$ of an experiment decreases, the gravitational 
attraction decreases as $L^{4}$.  It therefore becomes difficult at 
shorter distances to distinguish the Newtonian force between the test 
masses from the background forces.  Surface forces are especially 
troublesome, because they increase rapidly at small separations.  
The most sensitive experiments in the centimeter range to date have 
used torsion balances.  Klimchitskaya, et al., have proposed to use 
a torsional pendulum inside a rotating steel sphere with a 
non-concentric cavity, and they expect to improve the current limits 
by roughly an order of magnitude in the range above 100 $\mu$m 
\cite{Klimchitskaya}.

Another approach involves the use of small mechanical oscillators for 
the experimental test masses.  The experiment of Carugno, Fontana, 
Onofrio, and Rizzo measured the force between a small copper mass and 
a steel resonator driven at 30 and $1.66 \times 10^{4}$ Hz,
respectively \cite{Carugno}.  Using a heterodyne detection technique, 
this experiment was able to set limits on new macroscopic forces 
between, for example, $\alpha = 1 \times 10^{10}$ at $\lambda =
2\times 10^{-5}$ m and $\alpha = 8 \times 10^{7}$ at $\lambda = 
2\times 10^{-4}$ m.  These limits are roughly two orders of magnitude 
weaker than those shown in Fig.~\ref{fig:lim}, however, the authors hope to
improve their sensitivity by as much as five orders of magnitude using
fiber optic techniques.

Our own design is illustrated in Fig.~\ref{fig:expt}.  Small
mechanical resonators are used for both the source mass and force 
detector.  The source mass is driven at the resonant frequency of a 
detector torsional mode, nominally 1 kHz.  In the resonant mode of 
interest, the two rectangular sections of the detector counter-rotate.
The amplitude of motion is large for the smaller section immediately 
above the source mass, while the larger section is nearly stationary.
This double torsional arrangement provides a degree of isolation from 
the oscillator mounting, which decreases the mode damping
\cite{Kleiman}.

The principal advantage of this high-frequency resonant method is that
seismic and other vibrational backgrounds can be easily filtered in 
the kilohertz range. Vibration isolation is provided by supporting the
masses from stacks of brass disks connected by steel wires 
\cite{Fairbank,Taber}. Other backgrounds include electrostatic
forces, which are suppressed with a gold-plated shield between the 
masses, and magnetic forces due to both external fields and
contamination.  Background due to the Casimir effect is comparable to 
electromagnetic effects only for mass separations below 1 $\mu$m, 
roughly 1\% of the anticipated minimum gap.

If these backgrounds can be sufficiently reduced, the dominant source 
of noise will be thermal noise due to dissipation in the detector 
oscillator.  Based on this limitation, we calculate the expected 
sensitivity of the experiment to a new Yukawa-type interaction.

The Yukawa force between parallel plates separated by a distance
$d(t)$ is given by:
\begin{equation}
F_{Y}(t) = 2 \pi \alpha G \rho_{s} \rho_{d} A \lambda^{2} \exp{(-d(t)
  / \lambda)}[1 -\exp{(-t_{s} / \lambda)} ][1 - \exp{(-t_{d} / \lambda)}]
\label{eq:FYp}
\end{equation}
where $\rho_{s}$ and $\rho_{d}$ are the source and detector mass 
densities, and $t_{s}$ and $t_{d}$ are the thicknesses.  This would be
an exact expression if either plate had area $A$ and the other had 
infinite area, but for the real geometry there are small edge 
corrections.  Neglecting these effects, and assuming the geometry in 
Fig.~\ref{fig:expt}b, the torque exerted on the detector 
(the smaller rectangle only) via a Yukawa interaction with the source 
is given by:
\begin{equation}
N_{Y}(t) = \pi \alpha G \rho_{s} \rho_{d} A_{d} R \lambda^{2} \exp{(-d(t)
  / \lambda)}[1 -\exp{(-t_{s} / \lambda)} ][1 - \exp{(-t_{d} / \lambda)}],
\label{eq:NY}
\end{equation}
where $A_{d}$ is the area of the detector above the source mass, and
$R$ is the distance between the edge of the detector and the torsion axis.

The read-out signal from the transducer varies with the amplitude of
the applied torque.  The source mass is driven at the detector
resonant frequency $\omega_{0}$, and only the amplitude of the signal 
torque associated with this frequency is effective in driving the 
detector.  From Fourier analysis, this amplitude is:
\begin{equation}
|N_{Y}(\omega_{0})| = 2 \pi \alpha G \rho_{s} \rho_{d} A_{d} R
\lambda^{2} I_{1} (d_{0} / \lambda) \exp{(- \bar{d} / \lambda)}[1 - 
\exp{(-t_{s} / \lambda)} ][1 - \exp{(-t_{d} / \lambda)}],
\label{eq:NYw}
\end{equation}
where $I_{1} (d_{0} / \lambda)$ is the modified Bessel function.  Here
we have used $d(t) = \bar{d} + d_{0} \sin{\omega_{0} t}$, where $d_{0}
= (d_{max} - d_{min}) / 2$ is the amplitude of the source mass motion 
and $\bar{d} = (d_{max} + d_{min}) / 2$ is the average source-detector gap.

The rms thermal noise torque about the central axis is found from the 
mechanical Nyquist theorem to be:
\begin{equation}
N_{T} = \sqrt{\frac{4kT}{\tau} \left( \frac{mR^{2} \omega_{0}}{3 Q} \right)},
\label{eq:NT}
\end{equation}
where $m$ is the mass of the small rectangular section of the detector
oscillator, $Q$ is the detector quality factor, and $\tau$ is the 
measurement integration time.

The signal-to-noise ratio is then simply the ratio of
equations~\ref{eq:NYw} and~\ref{eq:NT}.  Setting the SNR equal to 
unity and solving for $\alpha$, we obtain the following result for 
the limit which the experiment may set on the Yukawa strength for each
value of the range: 
\begin{equation}
\alpha = \frac{2}{\pi \sqrt{3}} \frac{1}{G \rho_{s} \rho_{d} 
\lambda^{2} A_{d}} \left(\frac{1}{\epsilon}\right) \sqrt{\frac{kTm 
\omega_{0}}{Q \tau}}
\label{eq:alpha}
\end{equation}
The efficiency factor $\epsilon$, which is of order unity for an 
optimized geometry, is given by:
\begin{equation}
\epsilon = 2 I_{1} (d_{0} / \lambda) \exp{(- \bar{d} / \lambda)}[1 - 
\exp{(-t_{s} / \lambda)} ][1 - \exp{(-t_{d} / \lambda)}]
\label{eq:epsilon}
\end{equation}

The bold curve in Fig.~\ref{fig:lim}, representing the expected 
sensitivity of the experiment, is calculated using Eqs.~\ref{eq:alpha}
and~\ref{eq:epsilon} and the parameters in Table~\ref{tab:par}.  For 
each value of $\lambda$, the maximum separation $d_{max}$ is chosen so
as to maximize the efficiency $\epsilon$.  The maximum separation is 
allowed to vary between the anticipated upper limit of 1 mm and a
lower limit of 100 $\mu$m, below which the backgrounds are expected to
dominate.  From the limit curve in Fig.~\ref{fig:lim}, we expect the 
experiment to cover roughly 10 square decades of the ($\alpha,
\lambda$) parameter space previously unexplored below 1 mm.

\section{Conclusion}

Torsion balance experiments to the present have been sensitive to 
gravitational-strength forces down to millimeter distance scales.  
While very precise Casimir force measurements have recently been 
carried out and set new limits on effects in the range below 
100 $\mu$m, our calculations reveal that these experiments still allow 
for forces several million times stronger than gravity in this region.
Experiments of gravitational sensitivity in this range have become 
all the more imperative with several recent theoretical predictions 
of new forces, motivated especially by advances in string theory.  
The experiment described in this paper, which uses high-frequency 
mechanical oscillators for the test masses, is expected to reach 
gravitational strength down to approximately 50 $\mu$m, and to be
sensitive to much of the parameter space for these new predictions.

\section{Acknowledgments}

We would like to thank S. K. Lamoreaux for useful discussions of his 
Casimir force data, G. Giudice for an explanation of the superstring 
models, C. Dembowski for his work in our laboratory, and L. DiLella
and R. Onofrio for useful comments. We acknowledge T. Banks,
R. Sundrum and K. T. Mahanthappa for alerting us
to Refs.~\ref{ref:Banks},~\ref{ref:Sundrum}, and~\ref{ref:Arkani}, 
respectively.  This work is supported by NSF grant PHY97-22098 and 
the NSF-REU Program at the University of Colorado.  H. C. acknowledges
the support of the McNair Program at the University of Colorado.

% now the references. delete or change fake bibitem. delete next three
%   lines and directly read in your .bbl file if you use bibtex.

\clearpage
\begin{figure}[hp]
\caption{Parameter space for Yukawa-type forces in which the strength 
relative to gravity ($\alpha$) is plotted versus the range
($\lambda$).  Limit curves from experiments defining the excluded 
region are shown (bold lines) along with the anticipated sensitivity 
of the proposed experiment, as well as specific predictions for new 
long range forces (fine and dashed lines).}
\label{fig:lim}
\end{figure}

\begin{figure}[hp]
\caption{a) principal components of the apparatus; b) cross-section 
of central region showing dimensions used in calculation of
experimental sensitivity.}
\label{fig:expt}
\end{figure}
\clearpage
\begin{table}
\caption{Parameters used for experimental sensitivity calculation.}
\label{tab:par}
\begin{tabular}{lr}\hline
\multicolumn{1}{c}{\it Parameter} & \multicolumn{1}{c}{\it Value}\\ \hline
Signal Frequency, $\omega / 2 \pi$ & 1 kHz \\
Quality Factor, $Q$ & $1 \times 10^{6}$ \\
Minimum Source-Detector Gap, $d_{min}$ & $100\mu$m \\
Integration Time, $\tau$ & 1000 s \\
Detector Area, $A_{d}$ & 1 $ \mbox{cm}^{2}$ \\
Source Thickness, $t_{s}$ & $200\mu$m \\
Detector Thickness, $t_{d}$ & $200\mu$m \\
Source Density (W), $\rho_{s}$ & 19.3 $ \mbox{g/cm}^{3}$ \\
Detector Density (Si), $\rho_{d}$ & 2.3 $ \mbox{g/cm}^{3}$ \\
Temperature, $T$ & 4K \\
\end{tabular}
\end{table}

\clearpage
\noindent J. C. Long, Experimental Status of Gravitational-Strength Forces in
the Sub-Centimeter Regime, Figure 1.

\begin{figure}[hbtp]
\begin{center}
\epsfxsize=12cm \epsfbox{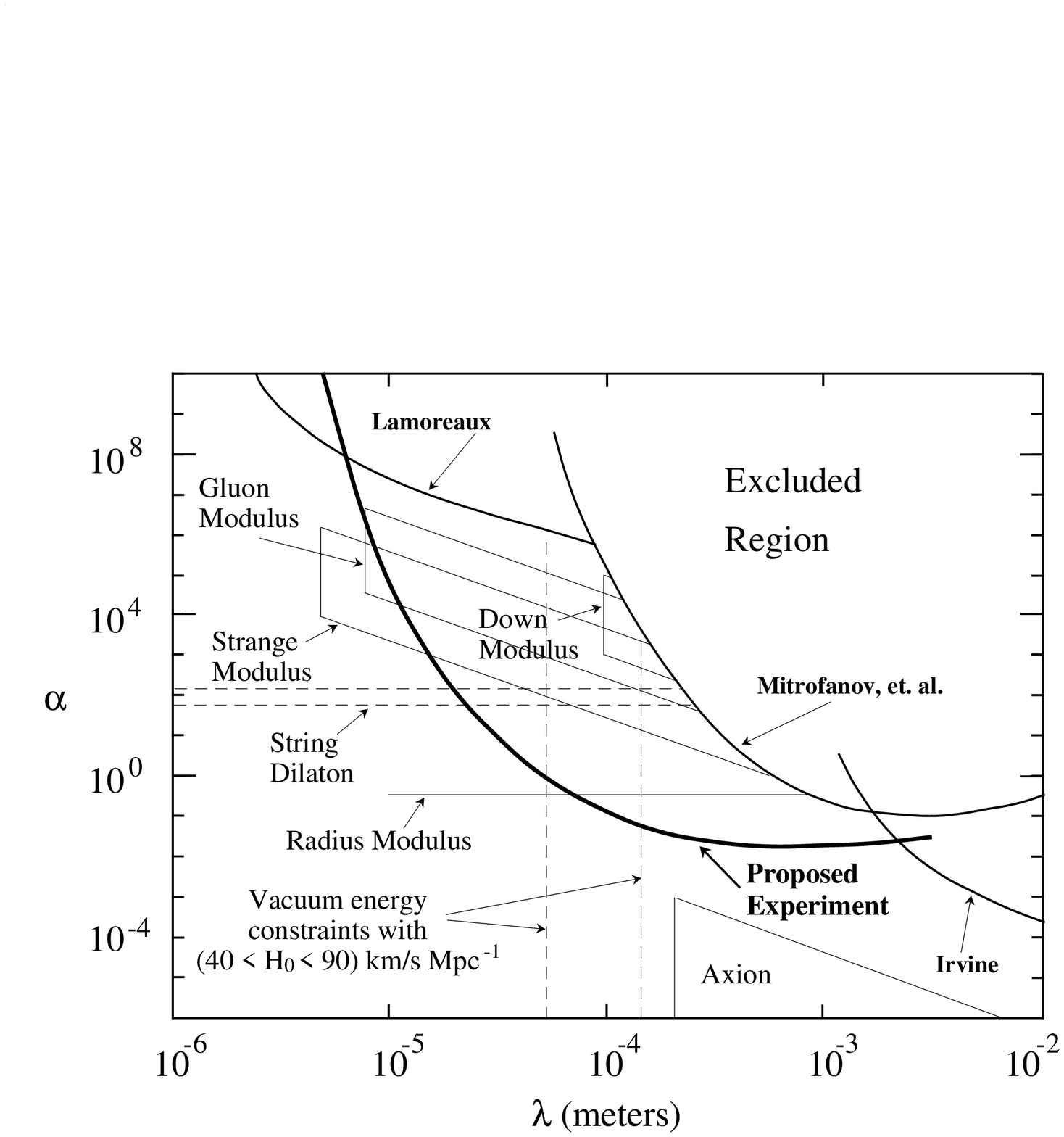}
\end{center}
\end{figure}

\clearpage
\noindent J. C. Long, Experimental Status of Gravitational-Strength Forces in
the Sub-Centimeter Regime, Figure 2.

\begin{figure}[hbtp]
\begin{center}
\epsfxsize=10cm \epsfbox{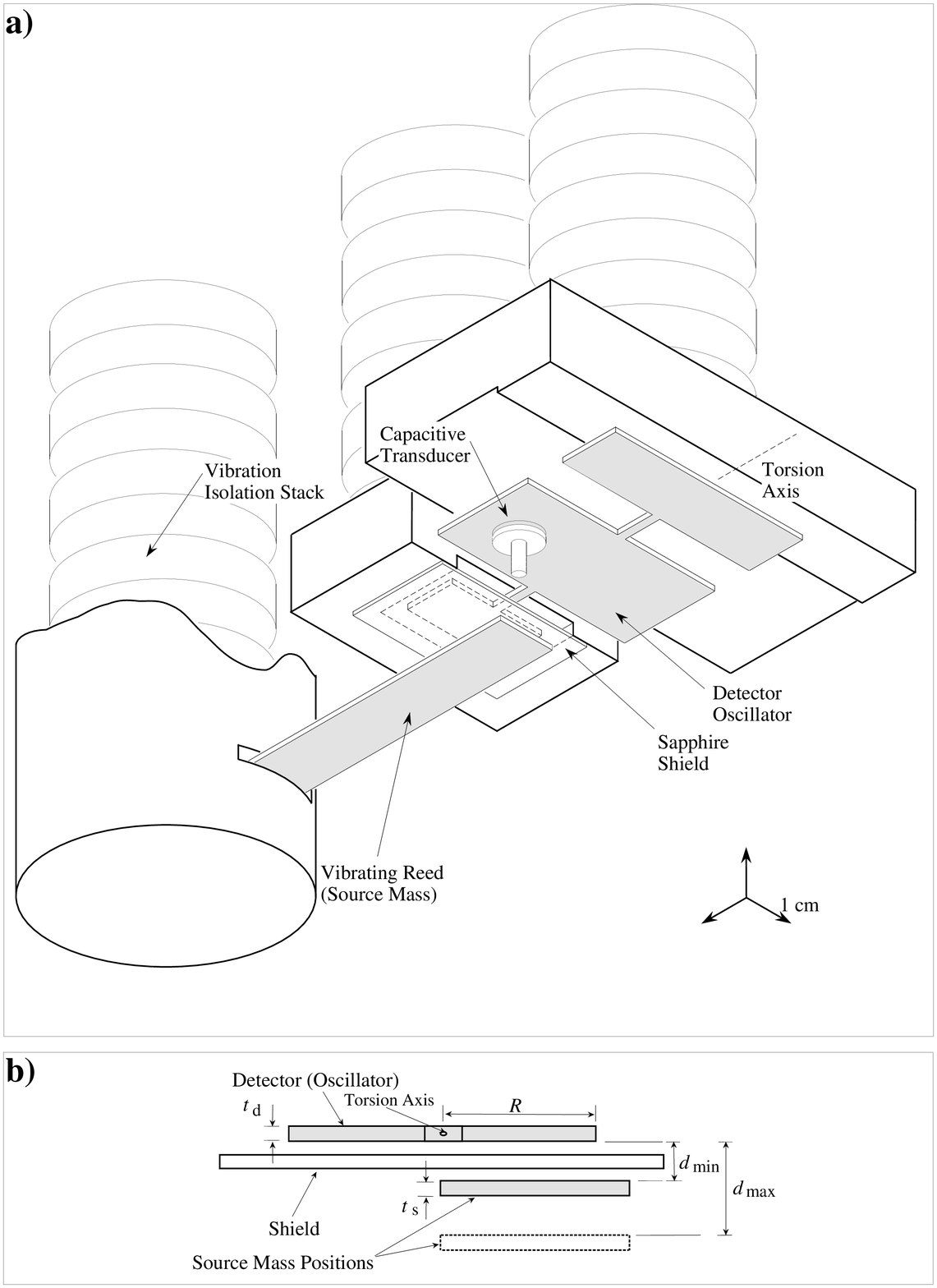}
\end{center}
\end{figure}

\begin{references}
\bibitem{Lamoreaux2} The actual form of $F_{Y}$ contains 17 
terms of the form \linebreak
$F_{Y} = 4 \pi G \rho_{1} \rho_{2} R \alpha \lambda^{3}
\exp{[-(a_{i} + a_{0}) / \lambda]} \exp{-d_{12} / \lambda}$.
This is the expression for the Yukawa force between a solid sphere and an
infinite plate or disk, where $G$ is Newton's Gravitational constant, 
$\rho_{1}$ and $\rho_{2}$ are mass densities, $R$ is the sphere
radius, $(a_{i} + a_{0})$ is the absolute disk-sphere separation, and
$d_{12}$ is an additional separation due to the thicknesses of the
Casimir plate materials.  Terms of this form are summed in a manner so
as to obtain, by superposition, the force between the composite lens
and disk used in the experiment.
\bibitem{Lamoreaux3} For negative values of $\alpha$, the limits are 
approximately the same as those of the Derjaguin experiment.  The
limits derived from the Mitrofanov and Irvine experiments for 
$\alpha < 0$, as reported in Refs.~\ref{ref:Hos} and~\ref{ref:Mit},
are weaker than the limits for $\alpha > 0$ only by small numerical
factors and are therefore not shown in Fig.~\ref{fig:lim}.
\bibitem{Lamoreaux4} Ref.~\ref{ref:Lam} includes summary and residual
plots which demonstrate a convincing signal for the Casimir effect
using the entire data set.  The data in these plots, however, are
adjusted for absolute disk-sphere separation and electromagnetic 
background as derived from the 3-component fit (Eq. 7 of
Ref.~\ref{ref:Lam}) to the measured force data.  Since the 3
components (Casimir, electromagnetic, and a constant term) do not
include a Yukawa term, which at current limits can be of the same
order as the largest of any of the other three, the summary data
cannot be used consistently to derive new force limits.  This limits 
our analysis to a re-fit of the small sub-sample of data presented in 
the measured (unadjusted) force plot.
\end{references}
\end{document}